\documentclass[11pt]{article}

\usepackage{pgfpages}
\usepackage{amsmath,amssymb,verbatim}
\usepackage{listings}
\usepackage{epsfig}
\usepackage[latin1]{inputenc}

\newcommand{\N}{I\!\!N}

\newcommand{\R}{I\!\!R}

\def\q{\quad}

\begin{document}

\tolerance=5000

\centerline{{\Large{\bf Multi-dimensional fractional wave equation 
}}}
\vspace{0.2cm}

\centerline{{\Large{\bf 
and 
some properties of its fundamental solution}}}

\vspace{0.3cm}
\centerline{{\bf Yuri Luchko}}
\vspace{0.3cm}

\centerline{{Department of Mathematics, Physics, and Chemistry}}

\centerline{{Beuth Technical University of Applied Sciences Berlin}}

\centerline{{Luxemburger Str. 10, 13353 Berlin,\ Germany}}

\centerline{{e-mail: luchko@beuth-hochschule.de}}

\vspace{0.2cm}

\begin{abstract}
\noindent
In this paper, a multi-dimensional fractional  wave equation that describes propagation of the damped waves is introduced and analyzed. In contrast to the fractional diffusion-wave equation, the fractional wave equation contains fractional derivatives of the same order $\alpha,\ 1\le \alpha \le 2$ both in space and in time. This feature is a decisive factor for inheriting some crucial characteristics of the wave equation like e.g. a constant phase velocity of the damped waves that are described by the fractional wave equation. Some new integral representations of the fundamental solution of the multi-dimensional wave equation are presented. In the one- and three-dimensional cases, the fundamental solution is obtained in explicit form in terms of elementary functions.   In the one-dimensional case, the fundamental solution is shown to be a spatial probability density function evolving in time. However, for the dimensions grater than one, the fundamental solution can be negative and therefore does not allow a probabilistic interpretation anymore.  To illustrate analytical findings, results of numerical calculations and numerous plots are presented.

\end{abstract}

\vspace{0.2cm}



\section{Introduction}

Fractional order differential equations have been successfully employed for
modeling of many different processes and systems in physics, chemistry, 	 engineering, medicine, biology, etc. (see e.g.  \cite{Dub}, \cite{FDL}, \cite{Her11}-\cite{Luc11_2}, \cite{LucPun11}, 
\cite{Mag04}-\cite{Mai_book},  \cite{Met04}, \cite{Uch08}) to mention only few of many recent publications). 

One of the most interesting and important topics in this field that has been considered by a number of researches since 1980's are models for anomalous transport processes  in form of time- and/or space-fractional advection-diffusion-wave equations.  Anomalous transport  represents a very natural way for description of time behavior of many complex systems that are characterized by a large diversity of elementary particles that participate
in the transport processes, by a strong interactions
between them, and by an anomalous evolution of
the whole system in time. In this paper, the focus will be on anomalous diffusion and wave propagation that are described by the fractional wave  equation that is also referred to as the neutral-fractional
diffusion equation; we do not
consider here the effects of the advection or the degradation
components of the general transport processes.

It is well known that whereas the diffusion equation describes a process, where a
disturbance of the initial conditions spreads infinitely fast, the propagation velocity of the disturbance is constant for the
wave equation. In this sense, the time-fractional diffusion-wave equation that is obtained from the diffusion equation by substituting the first derivative in time by the fractional derivative of order $\alpha,\ 1<\alpha <2$, interpolates between these two different behaviors. Namely, whereas a response of its fundamental solution to a localized
disturbance spreads infinitely fast,  its maximum  location disperses with a finite velocity $v(t,\alpha)$ that is determined by the formula (\cite{Fuj90}, \cite{LucMai}, \cite{LucMaiPov})  
\begin{equation}
\label{speed1}
v(t,\alpha) = C_\alpha t^{\frac{\alpha}{2}-1}.
\end{equation}
For $\alpha = 1$ (diffusion), the propagation velocity is equal to zero because of  $C_{1} = 0$, for $\alpha = 2$ (wave propagation) it remains constant, whereas for all intermediate values of $\alpha$ the propagation velocity of the maximum point depends on time $t$ and is a decreasing function that varies from $+\infty$ at time $t=0+$ 
to zero as $t\to +\infty$. This fact makes it difficult to interpret the  solutions to the time-fractional diffusion-wave equation as waves.

The fractional wave equation we deal with in this paper  contains fractional derivatives of the same order $\alpha,\ 1\le \alpha \le 2$ both in space and in time. The fractional derivative in time is interpreted in the Caputo sense whereas the space-fractional derivative is taken in form of the inverse operator to the Riesz potential (Riesz fractional derivative). It turns out that this feature of the fractional wave equation is a decisive factor for inheriting some crucial characteristics of the wave equation. In particular, we show that the phase velocity of the damped three-dimensional waves that are described by the fundamental solution of the three-dimensional fractional wave equation  is constant that depends just on the equation order $\alpha$. 

It is important to note that the anomalous transport models are usually first formulated in stochastic form in terms of the continuous time random walk 
processes. The 
time- and/or space-fractional differential equations are then derived from the stochastic models for a special choice of the jump probability density functions with the 
infinite first or/and second moments (see e.g. \cite{Kla08, Luc11_2, Met04}). The fractional wave equation can be obtained from the continuous time random walk 
processes, too. In \cite{Luc12}, the case of the waiting time probability density function and
the jump length probability density function with the same power low asymptotic behavior has been considered. Under some standard assumptions, the fractional wave equation cab be asymptocially derived  from the  continuous time random walk model mentioned above (see \cite{Luc12} for details). 

From the mathematical viewpoint, the fractional wave equation  was considered  for the first time in \cite{GorIskLuc}, where an explicit formula for the fundamental solution of the one-dimensional fractional wave equation was derived and where this equation was referred to as the neutral-fractional
diffusion equation. In \cite{MaiLucPag}, a one-dimensional space-time fractional diffusion-wave equation with the Riesz-Feller derivative of order $\alpha \in (0,2]$ and skewness $\theta$ and with the Caputo fractional derivative of order $\beta \in (0,2]$ was investigated in detail. A particular case of this equation called the neutral-fractional diffusion equation that for $\theta=0$ corresponds to our one-dimensional fractional wave equation  has been shortly mentioned in \cite{MaiLucPag}. In \cite{Met02}, a fundamental
solution to the one-dimensional neutral-fractional diffusion equation was deduced and analyzed in terms of the Fox
H-function. Finally, in \cite{Luc13}, the one-dimensional fractional wave equation was investigated in detail. Its fundamental solution $G_{\alpha,1}$  was derived in terms of elementary functions for all values of $\alpha,\ 1\le \alpha < 2$. Moreover, $G_{\alpha,1}$ was interpreted as a spatial probability density function evolving in time all whose moments of order less than $\alpha$ are finite. For the fundamental solution $G_{\alpha,1}$, both its maximum location and its maximum value were determined in closed form.  Finally, it was shown in \cite{Luc13}   that both the maximum and the gravity and "mass" centers of the fundamental solution $G_{\alpha,1}$ propagate with constant velocities like in the case of the wave equation, but in contrast to the wave equation ($\alpha = 2$) these velocities are different each to other for a fixed value of $\alpha,\ 1<\alpha <2$.  Moreover, the first, the second, and the Smith centrovelocities of the damped waves described by the one-dimensional fractional wave equation were shown to be constants that depend just on the equation order $\alpha$. 

In this paper, we deal with the multi-dimensional fractional wave equation with a special focus given to the most important three-dimensional case. The fundamental solution to this equation is a spherically symmetric function that possesses nice integral representations and can be even written down in explicit form in terms of elementary functions in the one- and three-dimensional cases. In contrast to the one-dimensional case, the fundamental solution for the multi-dimensional case cannot be interpreted as a probability density function and thus these equations cannot be employed for modeling of any diffusion processes. Instead, we show that the fundamental solution can be interpreted as a damped wave with the constant phase velocity that just depends on the order $\alpha$ of the fractional wave equation. 

The rest of the paper is organized as follows. In the 2nd section,  basic definitions, problem formulation, and some analytical results for the initial-value problems for the multi-dimensional fractional wave equation  are presented. In particular, the fundamental solutions $G_{\alpha,1}$ for the one-dimensional and $G_{\alpha,3}$ for the three-dimensional wave equations are derived in terms of elementary functions for all values of $\alpha,\ 1\le \alpha < 2$. Whereas $G_{\alpha,1}$ can be interpreted as a spatial probability density function evolving in time whose moments up to order $\alpha$ are finite, it is not true for the higher dimensions.  But the interpretation of the fundamental solutions as damped waves is still valid for the two- and three-dimensional cases. In particular, we show that the maximum location (phase) velocity of the fundamental solution $G_{\alpha,3}$ of the three-dimensional wave equation is a constant that depends just on the equation order $\alpha$.  To illustrate analytical findings, results of numerical calculations, numerous plots,  their physical interpretation and discussion are presented in the last section along with some conclusions and open problems for further research.

\section{Analysis of the fractional wave equation}

\subsection{Problem formulation}
In this paper, we deal with the initial-value
problem
\begin{equation}
\label{ic}
u(x,0)=   \varphi(x)\,,\ \ \frac{\partial u}{\partial t}(x,0)= 0, \ \ x\in \R^n
\end{equation}
for the multi-dimensional fractional wave equation 
\begin{equation}
\label{eq}
D^{\alpha}_t u(x,t)  \, = \,
-(-\Delta)^{\frac{\alpha}{2}} u(x,t), \q x \in \R^n\,, \q t \in {\R}_+,\ 1\le \alpha \le 2.
\end{equation}
In (\ref{eq}), $u=u(x,t),\ x \in \R^n\,, \ t \in {\R}_+$ is a real field variable,  $-(-\Delta)^{\frac{\alpha}{2}}$ is
the Riesz space-fractional
 derivative of order $\alpha $ that is defined below, and
 $D_t^\alpha$	is
 the Caputo time-fractional derivative of order
  $\alpha$: 
\begin{equation}
\label{fd}
(D^{\alpha}_t u)(t):= \left(I^{n-\alpha} \frac{\partial^n u}{\partial t^n} \right)(t), \ n-1<\alpha\le n,\ n\in \N
\end{equation}
$I^\alpha,\ \alpha \ge 0$ being the Riemann-Liouville fractional integral
$$
(I^{\alpha} u)(t):= 
\begin{cases}
\frac{1}{\Gamma(\alpha)} \int_0^{t} (t-\tau)^{\alpha -1} u(x,\tau)\, d\tau, \  \alpha >0, \\
u(x,t),\ \alpha =0
\end{cases}
$$
and $\Gamma$ the Euler Gamma function.
For $\alpha=n,\ n\in \N$, the Caputo fractional derivative coincides with the standard derivative of order $n$.  
 
All quantities in (\ref{eq}) are supposed to be dimensionless, so that the coefficient by 
the Riesz space-fractional
 derivative can be taken to be equal to one without loss of generality.  
 
In this paper, we are mostly interested in behavior and properties of the first 
fundamental solution (Green function) $G_{\alpha,n}$ 
of  the
equation (\ref{eq}), i.e. its solution  with the initial condition  $\varphi(x) = \prod_{i=1}^n \delta (x_i),\ x\in \R^n$, $\delta$ being  the Dirac delta function. 
  
For a sufficiently well-behaved function $f:\, \R^n \to \R$, the Riesz space-fractional
 derivative of order $\alpha, \ 0<\alpha \le 2$ is defined as a pseudo-differential operator with the symbol
$-|\kappa|^\alpha$ (\cite{Sai1}, \cite{Samko}):
\begin{equation}
\label{spfr}
({\cal F} -(-\Delta)^{\frac{\alpha}{2}} f)(\kappa) = -|\kappa|^\alpha ({\cal F} f)(\kappa),
\end{equation}
$({\cal F} f)(\kappa)$ being the Fourier transform of a function
$f$ at the point $\kappa \in \R^n$ that is defined by
\begin{equation}
\label{FTd}
({\cal F} f)(\kappa)\, =\, \hat f(\kappa) \, =\,   \int_{\R^n}
e^{i\kappa\cdot x}\, f(x)\, dx.
\end{equation}
For $0<\alpha < 2$, $\alpha\not = 1$, the Riesz space-fractional
 derivative can be represented as a hypersingular integral
$$
-(-\Delta)^{\frac{\alpha}{2}}f(x) =-\frac{1}{d_{n,l}(\alpha)}\int_{\R^n} \frac{(\Delta_h^l f)(x)}{|h|^{n+\alpha}}\, dh
$$
with the finite differences operator $(\Delta_h^l f)(x) = \sum_{k=0}^l (-1)^k \left( {l \atop k} \right) f(x-kh)$ and a suitable normalization constant $d_{n,l}(\alpha)$ (see \cite{Samko} for more details).  In particular, in the one-dimensional case we get the representation 
\begin{equation}
\label{Riesz}
-(-\Delta)^{\frac{\alpha}{2}}f(x) ={-1\over 2\Gamma(-\alpha)
\cos(\alpha\pi)}\int_0^\infty {f(x+\xi)-2f(\xi)+f(x-\xi)\over
\xi^{\alpha +1}}d\xi
\end{equation}
that is valid  for $0<\alpha < 2$, $\alpha\not = 1$.
For $\alpha =1$, the Riesz space-fractional derivative  can be interpreted in terms of the Hilbert transform (see e.g. \cite{GorMai})
$$
-(-\Delta)^{\frac{1}{2}}f(x) = -{1\over \pi} {d\over dx} \int_{-\infty}^{+\infty}
{f(\xi)\over x-\xi} d\xi,
$$
where the integral is understood in the sense of the Cauchy principal value.
In particular, in the one-dimensional case the fractional wave equation (\ref{eq}) with $\alpha = 1$ can be rewritten in the form
\begin{equation}
\label{eq-1}
\frac{\partial u}{\partial t}(x,t)  \, = \,
-{1\over \pi} {d\over dx} \int_{-\infty}^{+\infty}
{u(\xi,t)\over x-\xi} d\xi
\end{equation}
that is of course not the standard advection equation.

For $\alpha = 2$, equation (\ref{eq}) is reduced to the multi-dimensional wave equation. In what follows, we focus on the case $1\le \alpha < 2$ because the case $\alpha = 2$ (wave equation) is well studied in the literature.

\subsection{Fundamental solution of the fractional wave equation}

We start our analysis by applying the multi-dimensional Fourier integral transform (\ref{FTd}) with respect to the variable $x\in \R^n$ 
to the fractional wave equation (\ref{eq}) with $1<\alpha <2$ and to the initial conditions (\ref{ic}) with $\varphi (x) = \prod_{i=1}^n \delta(x_i)$. Using the definition of the Riesz fractional derivative, for the Fourier transform $\hat G_{\alpha,n}(\kappa,t)$ of the fundamental solution $G_{\alpha,n}(x,t)$ we get the initial-value problem
\begin{equation}
\label{4.4}
\left\{
\begin{array}{l}
\hat G_{\alpha,n}(\kappa,0)=1,    \\
\frac{\partial \hat G_{\alpha,n}}{\partial t}(\kappa,0)=0
\end{array} \right.
\end{equation}
for the fractional differential equation 
\begin{equation}
\label{4.3}
(D^{\alpha}_t \hat G_{\alpha,n}(\kappa,t))(t) +|\kappa|^\alpha\hat G_{\alpha,n}(\kappa,t) =0.
\end{equation}
The unique solution of the problem
(\ref{4.4})-(\ref{4.3}) is given by the expression (see e.g. \cite{Luc99A})
\begin{equation}
\label{4.5}
\hat G_{\alpha,n}(\kappa,t)  = E_\alpha(-|\kappa|^\alpha t^\alpha)
\end{equation}
in terms of the Mittag-Leffler function 
\begin{equation}
\label{M-L}
E_\alpha(z) =\sum_{k=0}^\infty {z^k\over \Gamma(1+\alpha k)},\ \alpha>0.
\end{equation}

It follows from (\ref{4.5}) and the well known asymptotic formula 
\begin{equation}
\label{M-L-a}
E_\alpha(-x) = -\sum_{k=1}^{m} {(-x)^{-k}\over\Gamma(1-\alpha k)} \ 
+O(|x|^{-1-m}),\ m\in \N,\ x\to +\infty,\ 0<\alpha <2
\end{equation}
for the Mittag-Leffler function  that $\hat G_{\alpha,n}$ belongs to the functional space $L_1(\R^n)$ with respect to $\kappa$ for $1<\alpha <2$. Therefore we can apply the inverse Fourier transform and get the representation
\begin{equation}
\label{fGR}
G_{\alpha,n}(x,t)=\frac{1}{(2\pi)^n} \int_{\R^n}
e^{-i\kappa\cdot x}\, E_{\alpha}
(-|\kappa|^\alpha t^\alpha)\, d\kappa,\ x\in \R^n,\ t>0
\end{equation}
for the fundamental solution $G_{\alpha,n}$. 
But $E_{\alpha}
(-|\kappa|^\alpha t^\alpha)$ is a radial function (spherically symmetric function) in $\kappa$ and thus the known formula (see e.g. \cite{Samko})
\begin{equation}
\label{FT}
\frac{1}{(2\pi)^n} \int_{\R^n}
e^{-i\kappa\cdot x}\, \phi(|\kappa|)\, d\kappa\, = \,
\frac{|x|^{1-\frac{n}{2}}}{(2\pi)^{\frac{n}{2}}} \int_0^\infty \phi(\tau)\tau^{\frac{n}{2}}J_{\frac{n}{2} -1}(\tau |x|)\, d\tau
\end{equation}
for the inverse Fourier transform of the radial functions can be employed under the condition that the integral at the right hand side of the formula (\ref{FT}) is conditionally or absolutely convergent. In this formula, $J_\nu$ stays for the Bessel function with the index $\nu$. 

The representation (\ref{fGR}) can be thus rewritten in the form 
\begin{equation}
\label{fGR-1}
G_{\alpha,\, n}(x,t) = \frac{|x|^{1-\frac{n}{2}}}{(2\pi)^{\frac{n}{2}}} \int_0^\infty E_{\alpha}
(-\tau^\alpha t^\alpha)\, \tau^{\frac{n}{2}}J_{\frac{n}{2} -1}(\tau |x|)\, d\tau
\end{equation}
under the condition that the integral at the right hand side of the formula is conditionally or absolutely convergent. 

Before we start with a convergence investigation of the integral in (\ref{fGR-1}), let us separately consider the case $x=(0,0,\dots,0) = {\mathbf 0}$ that corresponds to the case $|x|=0$. It follows from (\ref{fGR}) that 
$$
G_{\alpha,n}({\mathbf 0},t)=\frac{1}{(2\pi)^n} \int_{\R^n}
 E_{\alpha}
(-|\kappa|^\alpha t^\alpha)\, d\kappa,\ t>0.
$$
Switching to the spherical coordinates, we can represent the last integral in the form
$$
\int_{\R^n}
 E_{\alpha}
(-|\kappa|^\alpha t^\alpha)\, d\kappa =
$$
$$
 \int_0^\infty\int_0^{\pi}\dots\int_0^{\pi} \int_0^{2\pi}  E_{\alpha}
(-r^\alpha t^\alpha)\, r^{n-1}\sin^{n-2}(\phi_1)\cdots \sin(\phi_{n-2})\, dr\,d\phi_1\dots d\phi_{n-1}.
$$
Because of the known formula (see e.g. \cite{Pru86})
$$
\int_{0}^\pi \sin^{\alpha -1}(\phi)\, d\phi = \sqrt{\pi} \frac{\Gamma\left(\frac{\alpha}{2}\right)}{\Gamma\left(\frac{\alpha+1}{2}\right)},
$$
we first get the expression
\begin{equation}
\label{gx=0}
G_{\alpha,n}({\mathbf 0},t) = \frac{1}{2^{n-1}\pi^{\frac{n}{2}} \Gamma\left(\frac{n}{2}\right)} \int_0^\infty E_{\alpha}
(-r^\alpha t^\alpha)\, r^{n-1}\, dr
\end{equation}
that can be interpreted as the Mellin integral transform of the Mittag-Leffler function (see e.g. \cite{LucKir}) and thus evaluated: 
$$
G_{\alpha,n}({\mathbf 0},t) = \frac{1}{\alpha t^n} \frac{1}{2^{n-1}\pi^{\frac{n}{2}} \Gamma\left(\frac{n}{2}\right)} \frac{\Gamma\left(\frac{n}{\alpha}\right)\Gamma\left(1-\frac{n}{\alpha}\right)}{\Gamma\left(1-n\right)}.
$$
Let us emphasize the convergence condition $0<n <\alpha$ for the right-hand side  integral in (\ref{gx=0}). For $1<\alpha < 2$, this condition means that the fundamental solution $G_{\alpha,n}({\mathbf 0},t)$ is finite just in the one-dimensional case $n=1$. Because the Gamma function $\Gamma(z)$ has a pole at the point $z=0$, in the one-dimensional case we get the formula
$$
G_{\alpha,1}(0,t) = 0
$$
that is in accordance with the results presented in \cite{Luc13}. In all other cases, the fundamental solution of the fractional wave equation (\ref{eq}) is not finite at the point $x=(0,0,\dots,0) = {\mathbf 0}$ for $t>0$ like the fundamental solution to the multi-dimensional wave equation.

For $x \not = {\mathbf 0}$, we employ now the asymptotics (\ref{M-L-a}) of the Mittag-Leffler function and the known asymptotics of the Bessel function
\begin{equation}
\label{bes1}
J_\nu(r) = O(r^\nu),\ \mbox{as} \ r \to 0,
\end{equation}
\begin{equation}
\label{bes2}
J_\nu(r) = \sqrt{\frac{2}{\pi r}}\cos(r-\pi(1+2\nu)/4) + O (r^{-\frac{3}{2}}),\ \mbox{as} \ r \to +\infty
\end{equation}
to derive the condition $n <2\alpha +1$ for the conditional and the condition $n < 2\alpha -1$ for the absolute convergence of the  integral in the right-hand side of the representation (\ref{fGR-1}). For $1 < \alpha < 2$, these conditions means that this integral  is conditionally convergent at least for $n = 1,2,3$ and absolutely convergent for $n = 1$. 

Thus for $1 < \alpha < 2$ the representation (\ref{fGR-1}) is valid at least for $n = 1,2,3$ and we deal with these cases in the further discussions. 

Taking into account the relation
$$
J_\nu(z) = \left(\frac{z}{2}\right)^\nu W_{1,\nu+1}\left(-\frac{z^2}{4}\right) 
$$
between the Bessel function $J_\nu$ and the Wright function $W_{\alpha,\beta}$ that is defined as the convergent series 
$$
W_{\alpha,\beta}(z) = \sum_{m=0}^\infty \frac{z^m}{m!\, \Gamma(\alpha m +\beta)}, \ \ \alpha > -1,\ \ \beta \in \R,
$$
we get the representation 
$$
G_{\alpha,\, n}(x,t) = \frac{2}{(4\pi)^{n/2}} \int_0^\infty \tau ^{n-1}\,  E_{\alpha}
(-\tau^\alpha t^\alpha)\, W_{1,n/2 -1 }\left(-\frac{1}{4}\tau^2|x|^2\right)\, d\tau
$$
of the fundamental solution $G_{\alpha,\, n}$ in terms of two most known and used special functions of fractional calculus -  the Mittag-Leffler function $E_\alpha$ and the Wright function $W_{\alpha,\beta}$.  

Now we derive another important representation of the fundamental solution $G_{\alpha,n}$ in terms of a Mellin-Barnes integral.

For $x \not = 0$, the right-hand side of the representation 
$$
G_{\alpha,\, n}(x,t) = (2\pi)^{-n/2} |x|^{1-n/2} \int_0^\infty \tau ^{n/2}\,  E_{\alpha}
(-\tau^\alpha t^\alpha)\, J_{n/2 -1}(\tau |x|)\, d\tau 
$$
can be interpreted as the Mellin convolution of the functions $ E_\alpha(-t^\alpha \tau^\alpha)$ and $\tau ^{-n/2-1} J_{n/2 -1} (1/\tau)$ at the point $y = 1/x$. 

Using  the Mellin integral transform technique (Mellin transforms of the Mittag-Leffler and the Bessel functions, some elementary properties of the Mellin transform, and the Mellin convolution theorem) we get
the representation ($x \not = 0$)
\begin{equation}
\label{MTR}
G_{\alpha,n}(x,t)=
\frac{1}{\alpha \pi^{n/2}|x|^n} 
{1\over 2\pi i} \int_{L}
\frac{\Gamma\left({s\over \alpha}\right)
\Gamma\left(1-{s\over \alpha}\right) \Gamma\left({n\over 2}-{s\over 2}\right)}{\Gamma(1-s)
2^s
\Gamma\left({s\over 2}\right)} \left({t \over |x|}\right)^{-s}
ds
\end{equation}
of the fundamental solution $G_{\alpha,n}$ in terms of the Mellin-Barnes integral (Fox H-function). For the details regarding evaluation of the improper integrals of the Mellin convolution type we refer the interested readers to the recent paper \cite{LucKir} or to the book \cite{Mar}.

Because of the factor 
$
\Gamma\left({n\over 2}-{s\over 2}\right),
$
the Mellin-Barnes representation of $G_{\alpha,n}$ has a different structure for the even and the odd dimensions. 

In particular, for $n = 3$ the known formula $\Gamma(1 +z) = z\, \Gamma(z)$ for the Gamma function leads to the relation
$$
\frac{\Gamma\left({s\over \alpha}\right)
\Gamma\left(1-{s\over \alpha}\right) \Gamma\left({3\over 2}-{s\over 2}\right)}{\Gamma(1-s)
2^s
\Gamma\left({s\over 2}\right)} = \left({1\over 2}-{s\over 2}\right) \frac{\Gamma\left({s\over \alpha}\right)
\Gamma\left(1-{s\over \alpha}\right) \Gamma\left({1\over 2}-{s\over 2}\right)}{\Gamma(1-s)
2^s
\Gamma\left({s\over 2}\right)}
$$
that connects the kernel of the Mellin-Barnes representation of $G_{\alpha,3}$ with the kernel of $G_{\alpha,1}$.
This formula along with some elementary properties of the Mellin integral transform (see \cite{LucKir} or \cite{Mar}) induces the relation
\begin{equation}
\label{1D3D-1}
G_{\alpha,\, 3}(x,t) = \frac{1}{2\pi|x|^2} \left( G_{\alpha, 1}(x,t) + t\frac{\partial}{\partial t} G_{\alpha, 1}(x,t)\right),\ \ x\not = 0
\end{equation}
between the three-dimensional fundamental solution $G_{\alpha,3}$ and the one-dimensional fundamental solution $G_{\alpha,1}$. In the next subsection, we deduce another relation between these two fundamental solutions via the spatial differentiation.

\subsection{Particular cases of the fundamental solution}

In the one-dimensional case ($n=1$), the representation (\ref{fGR-1}) can be rewritten in the form
\begin{equation}
\label{1D}
G_{\alpha,\, 1}(x,t)={1\over \pi} \int_{0}^{\infty} E_{\alpha}
(-\tau^\alpha t^\alpha)\, \cos(\tau |x|)
\, d\tau,\ x\in \R,\ t>0
\end{equation}
because of the formula
$$
J_{-1/2}(x) = \sqrt{\frac{2}{\pi z}} \cos(z).
$$
The formula (\ref{1D}) is well known; it was used in \cite{GorIskLuc} and then in more detailed form in \cite{Luc13} to get an explicit form of the fundamental solution in terms of elementary functions:
\begin{equation}
\label{Green1}
G_{\alpha,1}(x,t)=
{1\over \pi} {|x|^{\alpha -1} t^\alpha \sin(\pi \alpha/2)\over
t^{2\alpha}+2|x|^\alpha t^\alpha \cos(\pi\alpha/2)+|x|^{2\alpha}},\ t>0,\ x\in \R, \ 1\le \alpha <2.
\end{equation}

For $\alpha = 1$ (modified advection equation (\ref{eq-1})), the fundamental solution $G_{\alpha,1}$ is the well known Cauchy kernel
\begin{equation}
\label{a=1}
G_{1,1}(x,t)=
{1\over \pi}{t\over t^2+x^2}
\end{equation}
that is a spatial probability density function evolving in time. 

For $\alpha =2$ (wave equation), the Green function $G_{2,1}$ is known to be given by the formula
\begin{equation}
\label{a=2}
G_{2,1}(x,t) = \frac{1}{2}(\delta(x-t)+\delta(x+t)).
\end{equation}
For $1 < \alpha <2$, the Green function (\ref{Green1}) 
is a spatial probability density function evolving in time, too, as has been shown for the first time in \cite{Luc13}.  
This pdf possesses all finite moments up to the order $\alpha$. In particular, the mean value of $G_{\alpha,1}$ (its first moment) exists for all $\alpha > 1$ (we note that the Cauchy kernel does not possess a mean value). The moments of  the order $\beta,\ |\beta| <\alpha$ of $G_{\alpha,1}$ 
can be evaluated via the Mellin integral transform (see \cite{Luc13} for details):
\begin{equation}
\label{mom}
\int_0^\infty G_{\alpha,1}(r,t)r^\beta\, dr =  \frac{t^\beta }{\alpha} \frac{\sin(\pi \beta/2)}
{\sin(\pi\beta/\alpha)}.
\end{equation}
We mention here that in \cite{Luc13} the locations of extrema points, gravity and mass centers of $G_{\alpha,1}$, and location of its energy center were calculated in explicit form and plotted for different values of the equation order $\alpha$. 

Another interesting and important feature of the one-dimensional fundamental solution $G_{\alpha,1}$ is that along with its probabilistic interpretation it can be interpreted as a damped wave, too. In \cite{Luc13}, different velocities of this wave were calculated and plotted depending on $\alpha$. In particular, the propagation velocity of its maximum that can be interpreted as the phase velocity, the propagation velocity of its gravity center, the velocity of its "mass"-center, its pulse velocity, and three different kinds of its centrovelocities have been introduced and calculated.  It turned out that all these velocities are constant in time and depend just on the order $\alpha$ of the fractional wave equation. 

\begin{figure}
\begin{center}
\includegraphics[width=8cm]{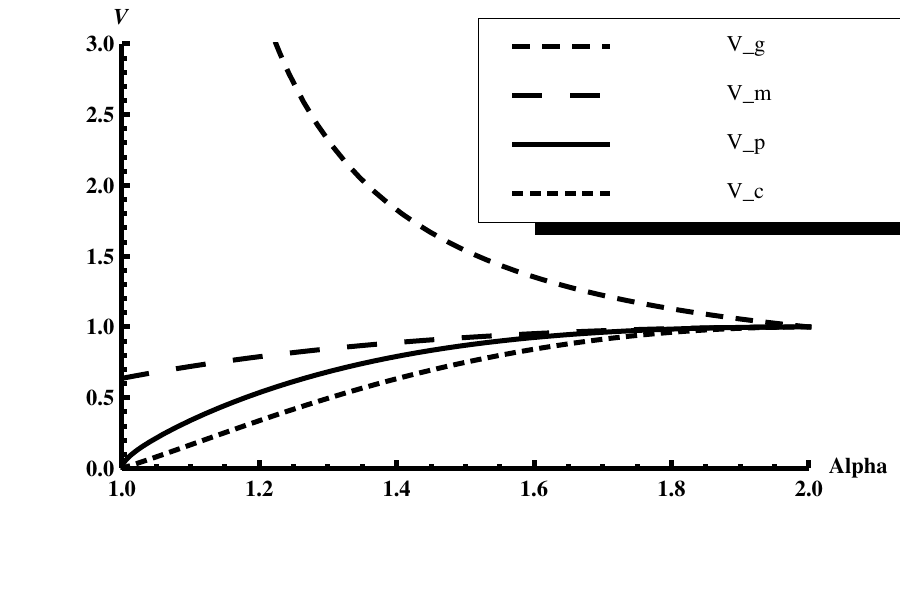}   
\caption{Plots of the gravity center velocity $v_g(\alpha)$,  the pulse velocity $v_m(\alpha)$, the phase velocity $v_p(\alpha)$, and the centrovelocity $v_c(\alpha)$ for $1\le \alpha \le 2$}
\label{fig-4}
\end{center}
\end{figure}

The plots of the propagation velocity $v_p$ of the maximum location of the fundamental solution $G_{\alpha,1}$ (phase velocity), the velocity $v_g$ of its gravity center, its pulse velocity $v_m$ and its centrovelocity $v_c$ are presented in Fig. \ref{fig-4}. As expected, $v_p=v_c=0$, $v_m = \frac{2}{\pi} \approx 0.64$ for $\alpha = 1$ (modified convection equation), and all velocities smoothly approach the value $1$ as $\alpha \to 2$ (wave equation). For $1<\alpha <2$, $v_p,\ v_m,$ and $v_c$ monotonously  increase whereas $v_g$ monotonously  decreases. It is interesting to note that  all velocities of $G_\alpha$ are nearly the same as those of the fundamental solution of the wave equation in a small neighborhood of the point $\alpha =2$. The velocity $v_g$ of the gravity center of $G_\alpha$ tends to $+\infty$ for $\alpha \to 1$ and $t>0$ (modified convection equation) because the first moment of the Cauchy kernel (\ref{a=1}) does not exist. It is interesting to note that for all $\alpha, 1<\alpha <2$ the velocities $v_p,\ v_g,\ v_m,\ v_c$ are different each to other and fulfill the inequalities $v_c(\alpha) <v_p(\alpha) <v_m(\alpha) <v_g(\alpha)$. For $\alpha = 2$, all velocities are equal to 1. 

In the two-dimensional case ($n = 2$), the representation (\ref{fGR-1}) has the form
\begin{equation}
\label{2D}
G_{\alpha,\, 2}(x,t) =  \frac{1}{2\pi} \int_0^\infty \tau \,  E_{\alpha}
(-\tau^\alpha t^\alpha)\, J_{0}(\tau |x|)\, d\tau,\ x\not = 0,\ t>0,
\end{equation}
where the Bessel function with the index $0$ can be represented as e.g. 
$$
J_0(z) = \int_0^\infty \cos(z\sin(\phi))\, d\phi.
$$
The representation (\ref{2D}) can be used for e.g. numerical evaluation of the fundamental solution $G_{\alpha,\, 2}$. The Mittag-Leffler function   $E_\alpha$ 
is evaluated by employing the algorithms suggested in \cite{Gor02} and the MATLAB-programs \cite{MC} that implement these
algorithms. Some results of the numerical evaluations of $G_{\alpha,\, 2}$ are presented in the next section. Say, the plot presented in Fig. \ref{fwe2Dfixedt} clearly shows that the fundamental solution $G_{\alpha,\, 2}$ is negative for some values of the variables $x$ and $t$ and therefore cannot be interpreted as a probability density function. 

Now we focus on the three-dimensional case. For $x\not = 0,\ t >0$, the representation (\ref{fGR-1}) takes the form
\begin{equation}
\label{3D_0}
G_{\alpha,\, 3}(x,t) = (2\pi)^{-3/2} |x|^{-1/2} \int_0^\infty \tau ^{3/2}\,  E_{\alpha}
(-\tau^\alpha t^\alpha)\, J_{1/2}(\tau |x|)\, d\tau
\end{equation}
that can be represented as
\begin{equation}
\label{3D}
G_{\alpha,\, 3}(x,t) = \frac{1}{2\pi^2|x|}  \int_{0}^{\infty} E_{\alpha}
(-\tau^\alpha t^\alpha)\, \tau \, \sin(\tau x)
\, d\tau
\end{equation}
because of the formula
$$
J_{1/2}(x) = \sqrt{\frac{2}{\pi z}} \sin(z).
$$
Comparing the representations (\ref{1D}) and (\ref{3D}),  
we easily get the relation ($x \not = 0$)
\begin{equation}
\label{1D3D-2}
G_{\alpha,\, 3}(x,t) = -\frac{1}{2\pi|x|} \frac{\partial}{\partial |x|} G_{\alpha, 1}(x,t)
\end{equation}
between the fundamental solutions of the one-dimensional and three-dimensional fractional wave equations. It is worth to mention that the same relation has been deduced in \cite{Han02} for the time-space-fractional diffusion equation with the time derivative of order $\alpha,\ 0<\alpha <1$ and the space derivative of order $\beta,\ 0<\beta<2$, $\beta \not = 1$ by using a different method.   

The representation (\ref{Green1}) of the fundamental solution $G_{\alpha,1}$ along with the relation (\ref{1D3D-1}) or the relation (\ref{1D3D-2}) 
between the fundamental solutions $G_{\alpha,1}$ and $G_{\alpha,3}$  leads to a closed form formula for fundamental solution of the  three-dimensional fractional wave equation  ($x\not = 0,\ t>0$):
\begin{equation}
\label{Green3}
G_{\alpha,\, 3}(x,t) = \frac{\sin(\pi \alpha/2)}{2\pi^2} \frac{ \left( -(\alpha -1)t^{2\alpha} + 2|x|^\alpha t^\alpha \cos(\pi\alpha/2) + (1+\alpha)|x|^{2\alpha}\right)}{
|x|^{3-\alpha} t^{-\alpha} \left(t^{2\alpha}+2|x|^\alpha t^\alpha \cos(\pi\alpha/2)+|x|^{2\alpha}\right)^2}.
\end{equation}
It follows from this formula that $G_{\alpha,3}$ is not a pdf because it is not everywhere nonnegative. More precisely,   the following inequalities are valid with $|x| = r >0$:
$$
G_{\alpha,\, 3}(r,t) < 0 \ \ \mbox{for} \ \ r < r^*(\alpha, t),
$$
$$
G_{\alpha,\, 3}(r,t) = 0 \ \ \mbox{for} \ \ r = r^*(\alpha, t),
$$
$$
G_{\alpha,\, 3}(r,t) > 0 \ \ \mbox{for} \ \ r > r^*(\alpha, t),
$$
where
\begin{equation}
\label{maxloc}
r^*(\alpha, t) = z_\alpha\, t,\ \ z_\alpha = \left(\frac{-\cos(\pi\alpha/2) + \sqrt{\alpha^2 - \sin^2(\pi\alpha/2)}}{\alpha +1}\right)^{\frac{1}{\alpha}}.
\end{equation}
Because of the relation (\ref{1D3D-2}) between the fundamental solutions $G_{\alpha,3}$ and $G_{\alpha,1}$,
the point $ r^*(\alpha, t) = z_\alpha\, t$
is the only maximum location of the fundamental solution $G_{\alpha, 1}(r,t)$ with $r>0,\ t>0$. In \cite{Luc13}, explicit formulas for  the location of the maximum point of $G_{\alpha,1}$, its maximum values and the monotonicity intervals were derived. The inequalities for the fundamental solution $G_{\alpha,3}$  follow from these results. 

From the closed form formula (\ref{Green3}), 
the asymptotics of $G_{\alpha,\, 3}(r,t),\ r = |x|>0$ with the fixed $t,\ t >0$ and $\alpha, 1 < \alpha < 2$ can be easily obtained:
$$
G_{\alpha,3}(r,t)= O(r^{\alpha -3}),\ \ r \to 0,
$$ 
$$
G_{\alpha,3}(r,t)= O(r^{-\alpha -3}),\ \ r \to +\infty.
$$ 
Thus the integral ("moment" of $G_{\alpha,3}(r,t)$ of the order $\beta$)
$$
I_{\alpha,\beta}(t) = \int_0^\infty r^\beta G_{\alpha,3}(r,t)\, dr
$$
exists for
$$
2 - \alpha <\beta < 2+\alpha,
$$
i.e., at least for
$$
1\le \beta \le 3 \ \ \mbox{if}\ \  1 < \alpha < 2. 
$$
Employing the the Mellin-Barnes representation of $G_{\alpha,3}$ ($x \not = 0$) in form
$$
G_{\alpha,3}(x,t)=
\frac{1}{\alpha \pi^{3/2}|x|^3} 
{1\over 2\pi i} \int_{L}
\frac{\Gamma\left({s\over \alpha}\right)
\Gamma\left(1-{s\over \alpha}\right) \Gamma\left({3\over 2}-{s\over 2}\right)}{\Gamma(1-s)
2^s
\Gamma\left({s\over 2}\right)} \left({t \over |x|}\right)^{-s}
ds
$$ 
that is a particular case of the formula (\ref{MTR}) and that 
can be interpreted as the inverse Mellin integral transform of its kernel,  we get an explicit formula for the Mellin integral transform of $G_{\alpha,3}$ and thus its "moments" of the order $\beta,\ 1\le \beta \le 3$: 
$$
I_{\alpha,\beta}(t) = \frac{(2t)^{\beta -2}}{\alpha \pi^{3/2}}
\frac{\Gamma\left({2-\beta\over \alpha}\right)
\Gamma\left(1-{2-\beta \over \alpha}\right) \Gamma\left({\beta +1 \over 2}\right)}
{\Gamma(\beta -1)
\Gamma\left({2-\beta \over 2}\right)}.
$$
Using the known properties of the Gamma function, a simpler  representation for the "moments" in form
$$
I_{\alpha,\beta}(t) = \frac{t^{\beta -2}(\beta -1)}{2 \alpha \pi} \frac{\sin\left( \frac{\pi \ \beta}{2}\right)}{\sin\left( \frac{\pi \ (2-\beta)}{\alpha}\right)}
$$
can be derived. In particular, 
we get the following important particular cases:

1) $\beta = 1$ ("`mean value"'):
$$
I_{\alpha,1}(t) \equiv 0,\ \ \mbox{for all} \ \ 1<\alpha<2,\ t>0,
$$

2) $\beta = 2$ (2nd "moment"):
$$
I_{\alpha,2}(t) \equiv \frac{1}{4\pi},\ \ \mbox{for all} \ \ 1<\alpha<2,\ t>0,
$$

3) $\beta = 3$ (3rd "moment"):
$$
I_{\alpha,3}(t) \equiv \frac{t}{\alpha \pi \sin(\pi/\alpha)},\ \ \mbox{for all} \ \ 1<\alpha<2,\ t>0.
$$

\subsection{Fundamental solution as a damped wave}

It follows from the Mellin-Barnes representation (\ref{MTR}) 
that $G_{\alpha,n}(x,t)$ can be represented via an auxiliary function that depends on a single argument:
\begin{equation}
\label{GL}
G_{\alpha,n}(x,t) = G_{\alpha,n}(|x|,t) = G_{\alpha,n}(r,t) = {r^{-n}}\, L_{\alpha,n}\left( \frac{r}{t}\right), \ r > 0 
\end{equation}
with 
\begin{equation}
\label{L}
L_{\alpha,n}\left( r \right) = \frac{1}{\alpha \pi^{n/2}} 
{1\over 2\pi i} \int_{L}
\frac{\Gamma\left({s\over \alpha}\right)
\Gamma\left(1-{s\over \alpha}\right) \Gamma\left({n\over 2}-{s\over 2}\right)}{\Gamma(1-s)
2^s
\Gamma\left({s\over 2}\right)} \left(r\right)^{s}
ds.
\end{equation}
Let the function ${r^{-n}}\, L_{\alpha,n}\left( r \right)$ attain its local maximum or minimum at  the point $r_* = c(\alpha,n)$.  
Then the fundamental solution $G_{\alpha,n}$ can be represented as
$$
G_{\alpha,n}(r,t) = t^{-n} \left( \frac{r}{t}\right)^{-n}\, L_{\alpha,n}\left( \frac{r}{t}\right)
$$
and therefore for a fixed $t>0$ $G_{\alpha,n}(r,t)$ attains its local maximum or minimum at the point
\begin{equation}
\label{maxlocn}
r_{*}(t,\alpha,n) = r_{*} \, t = c(\alpha,n)\, t.
\end{equation}
For $n=1$, the function ${r^{-n}}\, L_{\alpha,n}\left( r \right)$ is connected with the stable unimodal distributions and is unimodal, i.e., it has only one maximum point that means that the fundamental  solution $G_{\alpha,1}$ of the one-dimensional fractional wave equation has only one maximum point (see \cite{Luc13} for details). In the two-dimensional case, the fundamental  solution $G_{\alpha,2}$  has many (probably infinitely many) local minima and maxima as can be seen on the plot of 
Fig. \ref{fwe2Dfixedt}. Finally, the fundamental solution $G_{\alpha,3}$ has only one maximum point as can be seen on the plots of Fig. \ref{fweTfixedA=15} and is proved by direct calculations based on the closed form formula (\ref{Green3}). 

The value $G_{\alpha,n}^*(t)$ of the fundamental solution $G_{\alpha,n}(r,t)$ at a local minimum or maximum point
given by (\ref{maxlocn}) is equal to 
$$
G_{\alpha,n}^*(t) = (c(\alpha,n)\, t)^{-n}  L_{\alpha,n}\left( c(\alpha,n)\right) = \frac{1}{t^n}\, (c(\alpha,n))^{-n}  L_{\alpha,n}\left( c(\alpha,n)\right),
$$
where the function $L_{\alpha,n}$ is defined as in (\ref{GL}) and (\ref{L})
and thus the product
$$
\left(r_{*}(t,\alpha,n)\right)^n\, G_{\alpha,n}^*(t) = L_{\alpha,n}\left( c(\alpha,n)\right)
$$
is time independent. In particular, for $n=1$ we get the relation
$$
r_{*}(t,\alpha,1)\, G_{\alpha,1}^*(t) = L_{\alpha,1}\left( c(\alpha,1)\right),
$$
i.e., the product of the maximum location and the maximum value of the fundamental solution $G_{\alpha,1}$ is time independent and depends just on $\alpha$. For the numerical calculations and plots of the maximum location, maximum value and their product we refer the interested reader to \cite{Luc13}.

\section{Plots, discussions, and open problems}

In this section, some results of numerical evaluations of the fundamental solutions $G_{\alpha,n},\ n=1,2,3$, their plots, discussions, and open problems are presented. 

\subsection{Plots and discussions}
\begin{figure}
	\centering
		\includegraphics[width=8cm]{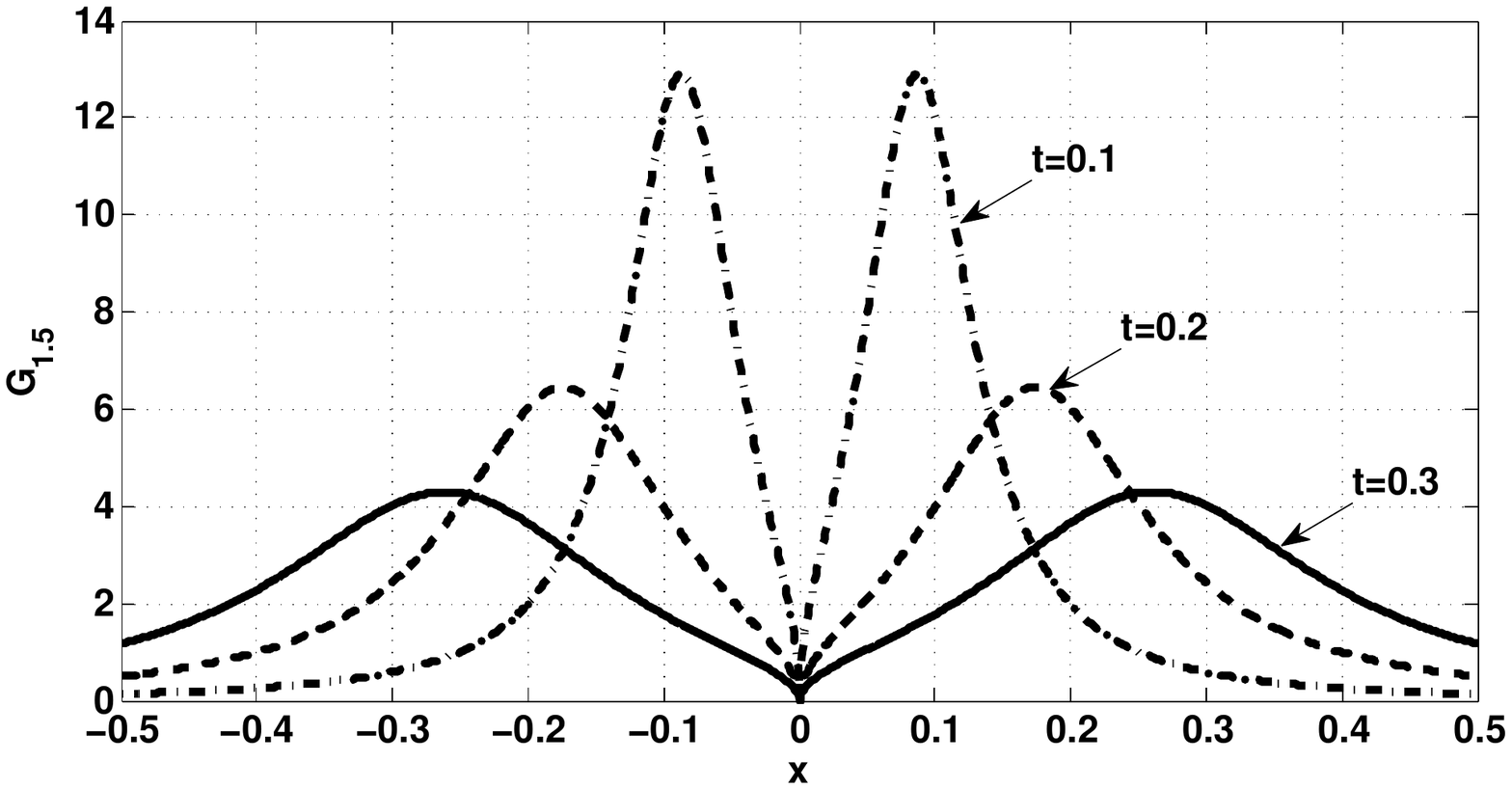}
		\caption{Plots of $G_{\alpha, 1}(x,t)$ with $\alpha = 1.5$ and for  different values of $t$}
		\label{2Dplot}
\end{figure}
\begin{figure}
\begin{center}
\includegraphics[width=8cm]{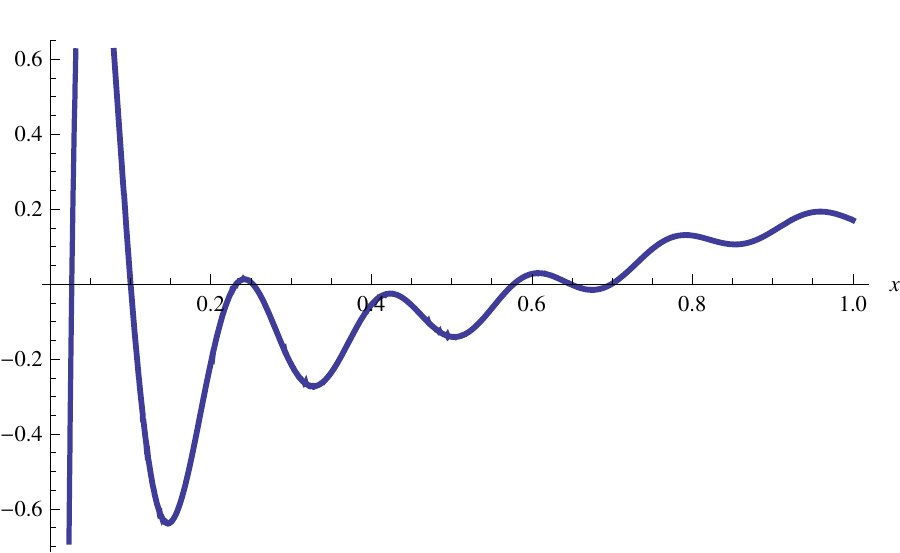}    
\caption{Plots of $G_{\alpha,2}(r,t),\ r=|x|>0$ with $\alpha=1.5$ at the time instant $t=1$ }
\label{fwe2Dfixedt}
\end{center}
\end{figure}
We start with some plots of the fundamental solution $G_{\alpha,1}$ that can be easily obtained using the closed form formula (\ref{Green1}) and are shown in Fig. \ref{2Dplot}. 
For more results regarding $G_{\alpha,1}$, its probabilistic and wave propagation interpretations we refer to \cite{Luc13}. 

On Fig. \ref{fwe2Dfixedt}, a plot of the fundamental solution $G_{\alpha,2}$ of the two-dimensional fractional wave equation is presented. To produce the plot, we employed the integral representation (\ref{2D}) and the MATLAB-programs \cite{MC}  for numerical evaluation of the Mittag-Leffler function $E_\alpha$ that implement the
algorithms suggested in \cite{Gor02}. An important information that can be read from the plot on Fig. \ref{fwe2Dfixedt} is that the fundamental solution $G_{\alpha,2}$  is negative for some values of the variables $x$ and $t$ and therefore cannot be interpreted as a probability density function. Another important point is that $G_{\alpha,2}$ is not unimodal and has many (probably infinitely many) local minimum and maximum points. 
\begin{figure}
\begin{center} 
\includegraphics[width=8cm]{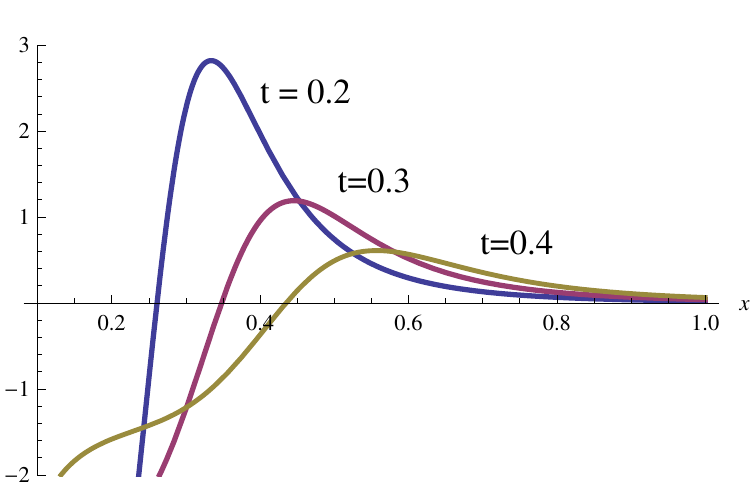}    
\caption{Plots of $G_{\alpha,3}(r,t)$ with $\alpha=1.5$ at the time instants $t=0.2,\ 0.3,\ 0,4$ }
\label{fweTfixedA=15}
\end{center}
\end{figure}
\begin{figure}
\begin{center}
\includegraphics[width=8cm]{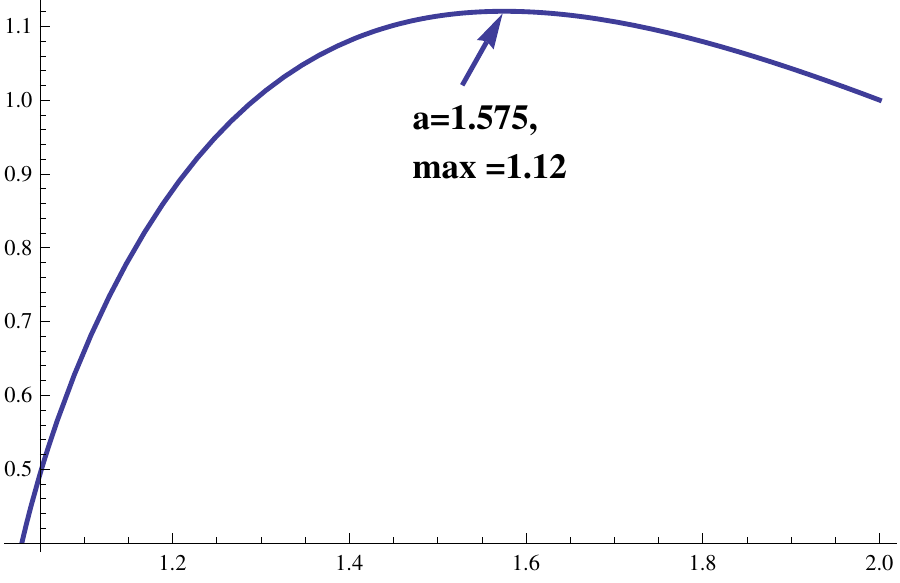}    
\caption{Plot of the phase velocity $v_p(\alpha)$ of $G_{\alpha,3}(r,t),\ r = |x|$ }
\label{fig_2}
\end{center}
\end{figure}
\begin{figure}
\begin{center}
\includegraphics[width=8cm]{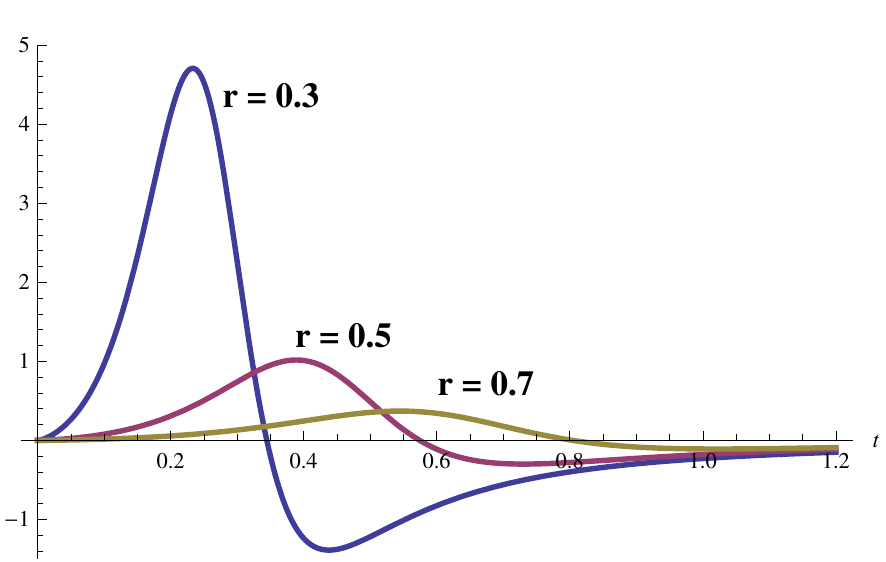}    
\caption{Plots of the fundamental solution $G_{\alpha,3}(r,t),\ r = |x|>0$ with $\alpha=1.5$ at the points $r=0.3,\ 0.5,\ 0,7$ }
\label{fweXfixedA=15}
\end{center}
\end{figure}
On the next Fig. \ref{fweTfixedA=15}, some plots of the fundamental solution $G_{\alpha,3}$ of the three-dimensional fractional wave equation are presented.

As we can see on the plots, for a fixed value of $t$, the fundamental solution $G_{\alpha,3}(r,t)$ has only one maximum point that depends on the time instant $t$ (and of course on the value of the parameter $\alpha$).
 
The velocity of the maximum location of $G_{\alpha,3}(r,t)$ (its phase velocity) can be calculated based on the formula (\ref{maxlocn}): 
$$
v_p(\alpha) := \frac{d r_{*}(t,\alpha,3)}{dt}  = 
\frac{d (c(\alpha,3)\, t)}{dt}   = c(\alpha,3),
$$
where $c(\alpha,3)$ is the location of the maximum point of the fundamental solution  $G_{\alpha,3}(r,t)$ at the time instant $t=1$. As we see, the phase velocity is time independent. It can be numerically calculated based on the closed form formula (\ref{Green3}) for the fundamental solution  $G_{\alpha,3}(r,t)$. The results of numerical calculations are presented in Fig. \ref{fig_2}. 

It is interesting to note that the phase velocity $v_p(\alpha)$ is not monotone in $\alpha$ and attains a maximum at the point $\alpha \approx 1.575$. In particular, it means that there exist infinitely many pairs $\alpha_1, \alpha_2,\ \alpha_1 \not = \alpha_2$ that fulfill the equality $v_p(\alpha_1) = v_p(\alpha_2)$, i.e., the corresponding damped waves propagate with the same  phase velocity.  

Finally, on Fig. \ref{fweXfixedA=15} some plots of the fundamental solution $G_{\alpha,3}$ with the  fixed $\alpha = 1.5$ and $|x| = r$ are presented.

As we see on the plots, the profiles of the fundamental solution $G_{\alpha,3}$ look like those of the damped waves that justifies our interpretation of the equation (\ref{eq}) as the fractional wave equation.


\subsection{Open problems}

In this paper, we discussed different representations and properties of the fundamental solution $G_{\alpha,n}$ of the fractional wave equation 
and interpreted it as a damped wave. In particular, the
maximum location of the fundamental solution and its velocity were determined. The maximum location velocity that can be interpreted as the phase velocity was shown to be a constant that depends just on the order $\alpha$ of the equation. 

Among open questions for further research,  investigation of other velocities of the damped waves that are described by the fractional wave equation like  the velocity of the mass-center, the pulse velocity, the group velocity, the centrovelocities, etc. should be first mentioned. Of course, the fractional wave equations with fractional derivatives of other types and  with the non-constant coefficients should be investigated, too.  In particular, the fractional wave equation with the Riesz-Feller derivative of order $\alpha \in (0,2]$ and skewness $\theta$ would be for sure an interesting research object. We note here that in \cite{MaiLucPag} an one-dimensional space-time fractional diffusion-wave equation with the Caputo derivative of order $\beta \in (0,2]$ in time and the Riesz-Feller derivative of order $\alpha \in (0,2]$ and skewness $\theta$ in space has been investigated in detail. A particular case of this equation called neutral-fractional diffusion equation that for $\theta=0$ corresponds to our one-dimensional fractional wave equation  was introduced in \cite{MaiLucPag}. It would be interesting to consider a multi-dimensional analog of the neutral-fractional diffusion equation that corresponds to the one-dimensional case considered in \cite{MaiLucPag}. 

Another potentially interesting research direction would be to employ the fractional wave equation introduced in this paper in the theory of the fractional Schrödinger equation (see e.g. \cite{Bader} or \cite{Luc13_1} for more details). Until now, different kinds of the space-, time, and space-time-fractional Schrödinger equations were introduced and analyzed, but not a fractional Schrödinger equation that corresponds to our fractional wave equation. 

Finally, we mention a very recent research field of fractional calculus that deals with the inverse problems for the fractional differential equations (see e.g. \cite{LucRun} and references there). It would be very interesting to consider some inverse problems for the multi-dimensional fractional wave equation introduced in this paper and their applications. 

\end{document}